\documentclass[aps, prl,
twocolumn, %showkeys, preprintnumbers,
amssymb,superscriptaddress]{revtex4}
\usepackage{graphicx, bm, amsmath, amsfonts,amssymb}

\usepackage{subfigure}

\def\v#1{\textbf{\emph{#1}}}

\def\up{\uparrow}
\def\down{\downarrow}
\def\dd{\mathrm{d}}

\begin{document}

%\preprint{}

\title{Emergence of $p+ip$ superconductivity in $2$D strongly correlated Dirac fermions}

\author{Zheng-Cheng Gu}
\affiliation{Perimeter Institute for Theoretical Physics, Waterloo, Ontario, N2L2Y5, Canada}

\author{Hong-Chen Jiang}
\affiliation{Stanford Institute for Materials and Energy Sciences, SLAC National Accelerator Laboratory, Menlo Park, CA 94025, USA}

\author{G. Baskaran}
\affiliation{The Institute of Mathematical Sciences, C.I.T. Campus, Chennai 600 113, India}
\affiliation{Perimeter Institute for Theoretical Physics, Waterloo, Ontario, N2L2Y5, Canada}

\date{\today}
\begin{abstract}
Searching for $p+ip$ superconducting(SC) state has become a fascinating subject in condensed matter physics, as a dream application awaiting in topological quantum computation. In this paper, we report a theoretical discovery of a $p+ip$ SC
ground state (coexisting with ferromagnetic order) in honeycomb lattice Hubbard model with infinite repulsive interaction at low doping($\delta< 0.2$), by using both the state-of-art Grassmann tensor product state(GTPS) approach and a quantum field theory approach. Our discovery suggests a new mechanism for $p+ip$ SC state in generic strongly correlated systems and opens a new door towards experimental realization. The $p+ip$ SC state has an instability towards a potential non-Fermi liquid with a large but finite $U$. However, a small Zeeman field term stabilizes the $p+ip$ SC state. Relevant realistic materials are also proposed.
\end{abstract}

\maketitle

\section*{Introduction}
Since the discovery of high-Tc Cuprates\cite{highTc}, the ground state phase diagram of the repulsive-$U$ Hubbard model in the strong coupling limit has become one of the most important open problems in condensed matter physics\cite{RVB,PatrickRVB}.
Infinite-$U$ Hubbard model is a simple looking but challenging many body problem that has defied exact understanding so far. While spin and charge degrees of freedom are manifestly decoupled, they might cooperate in surprising ways giving rise to novel many body states that are yet to be understood. In this paper, we present one such surprise, namely a Nagaoka state\cite{Nagaoka} nurturing a $p + ip$ superconductivity in $2$D honeycomb lattice infinite-$U$ Hubbard model at small but finite doping(its noninteracting counterpart is described by Dirac pockets), at the expense of only a tiny reduction of its own saturated ferromagnetism.

At exact half filling, all  low energy states are exhausted by the 2$^N$($N$ is the total number of sites) fold degenerate ground states generated by the localized free (dangling) spin-half moments.  Spins have no dynamics of their own. Upon doping,  holes hop and scramble spin configurations. In general, coherent delocalization of charges becomes difficult and kinetic energy gets frustrated, except when we have saturated ferromagnetism and a simple fermi sea as a ground state. While coherent delocalization of charges occurs through Bloch state formation, Pauli principle increases the kinetic energy, because only up spin bands are used. However, the real question is, are there other organizations of charge and spin degrees of freedom that reduce kinetic energy frustration and beat the Nagaoka state?

There are conflicting results for the stability of Nagaoka state in square lattice infinite-$U$ Hubbard model at small hole doping. An analysis by Wen and Doucot indicates\cite{WenNagaoka} that even for 2-holes the ground state is not fully spin polarized. There is also an argument given by Trugman\cite{TrugmanNagaoka} that in any finite system the lowest energy state cannot have bound holes.  Series expansion results also cast doubt on stability of Nagaoka ferromagnet at any finite doping\cite{seriesNagaoka}. Recent numerical calculations\cite{DMRGNagaoka} come to opposite conclusions. Therefore, further analysis of the intriguing situation for square lattice and other lattice geometries are called for.

Keeping the above in mind, we report a detailed state-of-art Grassmann tensor product state(GTPS) numerical calculation\cite{FrankfPEPS,ifPEPS,fermionicTPS,GPEPS,finitefPEPS,GuGTPS,Gumethod,honeycombtJ,squaretJ1,squaretJ2}, an effective theory and a physically motivated approximate microscopic analysis of the honeycomb lattice infinite-$U$ Hubbard model with hole doping $\delta = 0\sim0.2$.  We find a surprising result that Nagaoka ferromagnet can support and nurture a $p + ip$ superconducting state. It is a win-win situation in the sense it grows a $p + ip$ order parameter only at the expense of a tiny reduction of its saturated magnetism, and gains an additional stability of its own ferromagnetism. What we have found can be interpreted as an interesting ${\it zero ~ point ~ reduction}$  of saturation ferromagnetism, unknown in the case of standard quantum ferromagnets.  For realistic systems with large but finite $U$, the $p+ip$ SC state has an instability towards a potential non-Fermi liquid state; however, by applying a Zeeman field,
the $p+ip$ SC state can be stabilized.

\section*{Results}
\emph{Ground state energy and magnetization} -- It is well known that the Hubbard model with strong repulsive interaction can be effectively described by the $t-J$ model\cite{tJ}:
\begin{equation}
H_{t-J}=t\sum_{\langle ij\rangle,\sigma
}(\tilde{c}^\dagger_{i,\sigma}
\tilde{c}_{j,\sigma}+h.c.)+J\sum_{\langle ij \rangle}\left(\vec S_i
\cdot\vec S_j-\frac{1}{4}n_in_j\right),\nonumber
\end{equation}
where $\tilde{c}_{i,\sigma}\equiv c_{i,\sigma}(1-n_{i,\bar\sigma})$ is the electron annihilation operator defined in
the no-double-occupancy subspace and $J=4t^2/U$. It is obvious that the infinite-$U$ limit corresponds to the $J=0$ limit.
%The single hole problem for $J=0$ limit was solved by Nagaoka, however, at any finite doping, the ground state problem turns out to be extremely hard.
Here we use a translationally invariant GTPS variational ansatz to investigate the possbile ground state phase for $J=0$, which is specified by just two different Grassmann tensors $\textbf{T}_A,\textbf{T}_B$ on sublattices $A,B$ in each unit cell(see the method section for a detailed explanation). We then use the imaginary time evolution method\cite{ImagTPS,Gumethod} to update
the GTPS from a randomly initialized state. Finally we use the weighted
Grassmann-tensor-entanglement renormalization group (wGTERG)
method\cite{GuGTPS,Gumethod} to calculate physical quantities.
The total system size is up to $2\times 27^2$ sites and all
calculations are performed with periodic boundary conditions(PBC). The
largest virtual dimension of the GTPS considered is $14$. To ensure the
convergence of the wGTERG method, we keep $D_{cut}$ (defined in
Refs.~\cite{GuGTPS,Gumethod}) up to $160$ for $D=10,12$ and $196$ for $D=14$ , which
gives small relative errors for physical quantities of order $10^{-3}$.

\begin{figure}[t]
{\includegraphics[width=0.4\textwidth]{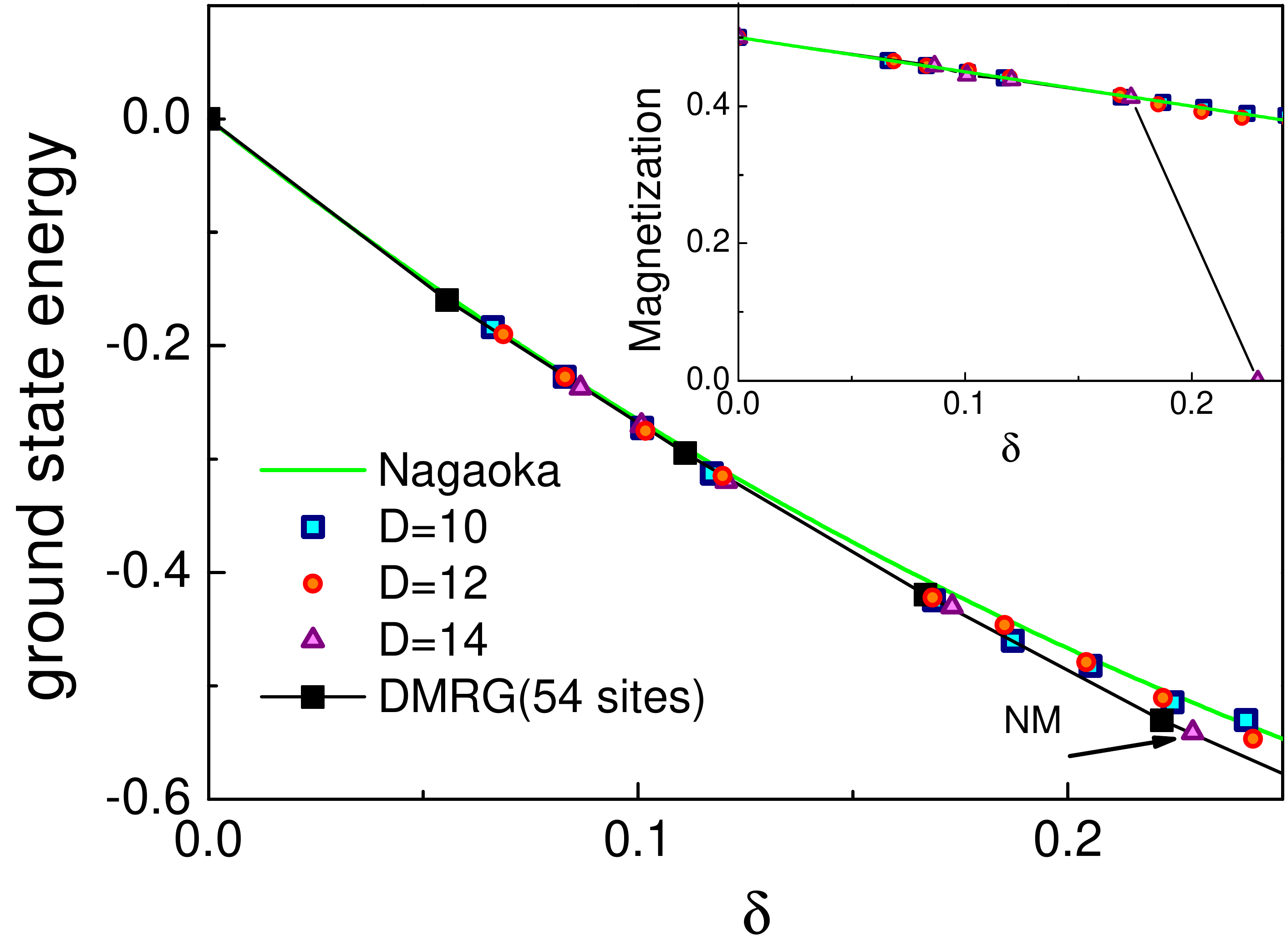}} \caption{(Color online)Ground
state energy as a function of doping. As a benchmark, we performed
DMRG calculation for a small cluster with $N=54$ sites under PBC.
Insert: FM magnetization as a function of
doping.}\label{fig:energy}
\end{figure}

As seen in Fig. \ref{fig:energy}, the ground state
energy shows a marked increase in $D$ dependence as
hole doping $\delta$ increases(a chemical potential term is added to control $\delta$). As a benchmark, we perform the density matrix renormalization group(DMRG)\cite{DMRG1992}
calculations for a small cluster with $N=54$($2\times 3\times 9$) sites under PBC.
DMRG is the only unbiased
method for frustrated systems that avoids the sign problem, but it is
restricted to relatively small systems, especially for PBC.
To ensure the convergence of the DMRG calculation with PBC,
we keep up to $8000$ states and make the truncation errors
less than $10^{-9}$ in our $N=54$ calculations. Up to $D=14$, we
find a systematic consistency for ground energy with both methods for $\delta < 0.25$.
We also compare with the Nagaoka state -- a fully polarized half metallic state(HMS), and find our ground state energy is significantly lower. This indicates the instability of HMS in honeycomb lattice infinite-$U$ Hubbard model at finite doping.

Despite of the instability of HMS, we find the ferromagnetic(FM) magnetization
$m=\sqrt{{\langle S^x_{i} \rangle}^2+{\langle S^y_{i} \rangle}^2+{\langle S^z_{i} \rangle}^2}$
is still extremely close to the fully polarized HMS at low doping. As seen in the insert of Fig. \ref{fig:energy}, up to $D=14$, we find $m/m_{\rm{max}}\sim 0.99 $ for $\delta<0.2$(e.g., $m/m_{\rm{max}}\sim 0.991$ for $\delta=0.1$) while $m=0$ for $\delta>0.2$, which is similar to recent DMRG results for square lattice infinite-$U$ Hubbard model\cite{DMRGNagaoka}. (It is clear that the non-vanishing $m$ for $D=10,12$ cases is due to a local minimum effect.)We also observe $\langle n_{i \in A} \rangle=\langle n_{j\in B} \rangle$ for $\delta<0.25$
and there is no commensurate charge density wave(CDW) order.

\begin{figure}[t]
{\includegraphics[width=0.4\textwidth]{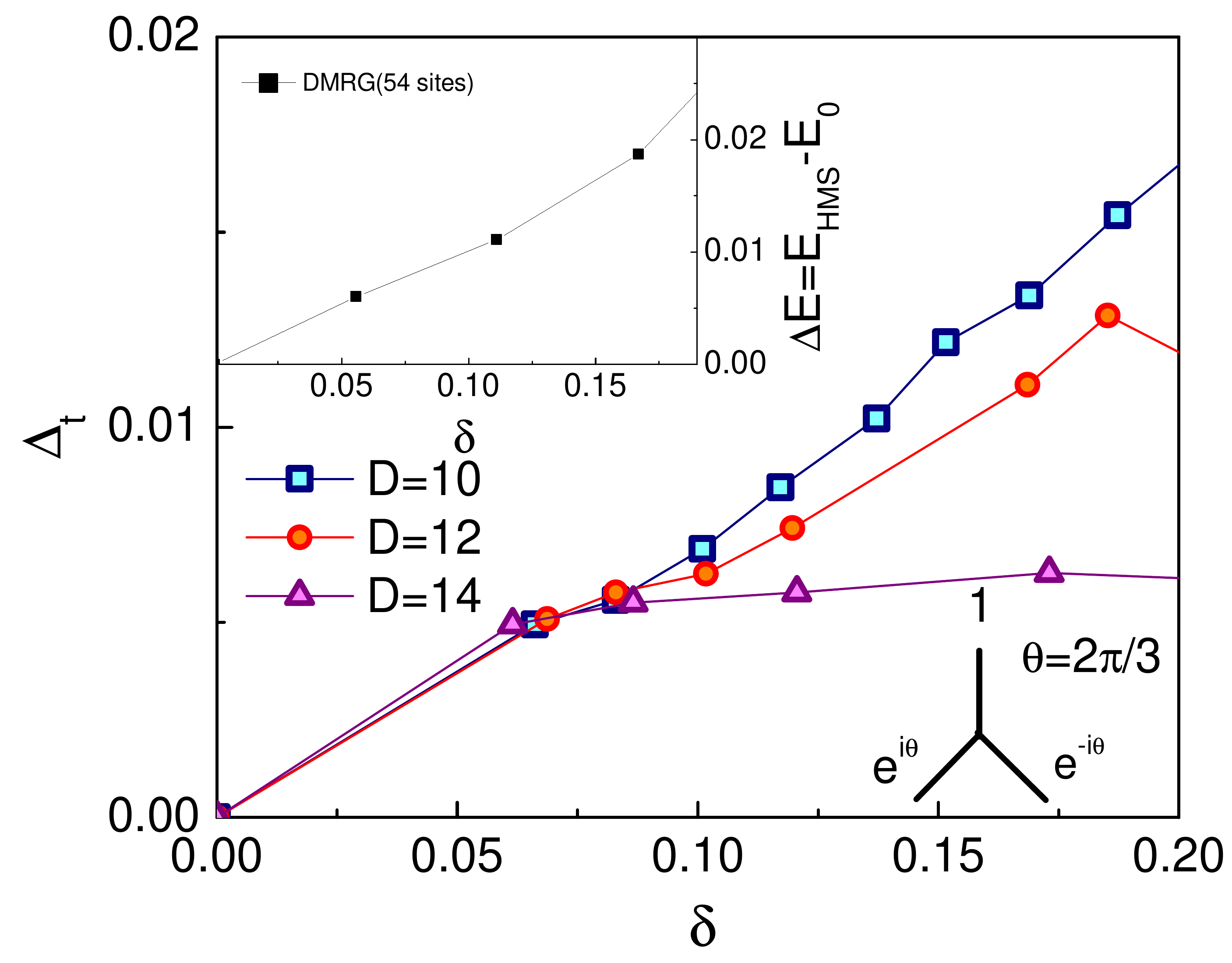}} \caption{(Color online)Triplet SC order
parameters as a function of doping. Insert: "condensation energy" $\Delta E$ as a function of
doping.}\label{fig:SC}
\end{figure}

\emph{The emergence of $p+ip$ superconductivity} --
Next we turn to the interesting
question of whether the doped honeycomb lattice infinite-$U$ Hubbard model
supports superconductivity or not, and if so, what
its pairing symmetry is. To answer this question, we calculate the real space
SC order parameters in the spin triplet channel
(the singlet channel vanishes due to the presence of FM order).
Since the triplet pairing order parameter has three independent
components $\vec\Delta_t = \frac{1}{\sqrt{2}}c_{i,\alpha} (i\sigma^y
\vec\sigma)_{\alpha\beta}
  c_{j,\beta}$($i \in A$, $j \in B$), we can define the amplitude of
triplet order parameter as $\Delta_t = \sqrt{\vec\Delta_t^* \cdot
\vec\Delta_t}$. Because we use a
chemical potential to control the hole concentration, the charge
$U(1)$ symmetry can be spontaneously broken in the variational
approach, which allows $\Delta_t$ to be measured directly rather
than through its two-point correlation function. As shown in the
main panel of Fig. \ref{fig:SC}, up to $\delta=0.2$ we find a
non-zero triplet SC order parameter for the whole region and $\Delta_t$ is almost $D$ independent for $\delta<0.1$.
By measuring the SC order parameters for the three inequivalent
nearest-neighbor bonds, we found
$\Delta_{t;a}^{x(y,z)}/\Delta_{t;b}^{x(y,z)}\simeq \Delta_{t;b}^{x(y,z)}/\Delta_{t;c}^{x(y,z)}\simeq \Delta_{t;c}^{x(y,z)}/\Delta_{t;a}^{x(y,z)}\simeq e^{i\theta}$
with $\theta=\frac{2\pi}{3}$(see Table \ref{phase}).
This pairing is consistent
with the $p+ip$ pairing symmetry. We note that the above result is quite nontrivial since we start with a
completely random state without pre-assuming SC order.
\begin{table}[h]
\begin{tabular}{|c|c|c|c|}
\hline
Doping & $\delta=0.069$ & $\delta=0.102$ & $\delta=0.168$  \\
\hline
$\Delta^{x,(y,z)}_{t;a}/\Delta^{x,(y,z)}_{t;b}$ & (-0.500,0.866)   &  (-0.500,0.866) & (-0.500,0.866) \\
$\Delta^{x,(y,z)}_{t;b}/\Delta^{x,(y,z)}_{t;c}$ & (-0.500,0.866)   &  (-0.500,0.866) & (-0.499,0.865) \\
$\Delta^{x,(y,z)}_{t;c}/\Delta^{x,(y,z)}_{t;a}$ & (-0.500,0.866)   &  (-0.500,0.866) & (-0.500,0.866) \\
\hline
\end{tabular}
\caption{Starting from a randomly initialized GTPS ansatz, we observed $\Delta_{t;a}^{x(y,z)}/\Delta_{t;b}^{x(y,z)}\simeq \Delta_{t;b}^{x(y,z)}/\Delta_{t;c}^{x(y,z)}\simeq \Delta_{t;c}^{x(y,z)}/\Delta_{t;a}^{x(y,z)}\simeq e^{\frac{2\pi i}{3}}=(-\frac{1}{2},\frac{\sqrt{3}}{2})$ for the (variational) ground state obtained from imaginary time evolution.(Here we use the data with inner dimension $D=12$ as an example.)}\label{phase}
\end{table}

We find
the SC and FM order coexist in the regime $0<\delta<0.2$.
 We further check the
internal spin direction of the triplet pairing vector $\vec\Delta_t$ and find it is
always perpendicular to the FM order parameter $\langle
\v S_i\rangle$. At larger doping close to
$\delta \sim 0.2$, the triplet order parameter has a very strong $D$
dependence. This indicates a possible phase transition around $\delta \sim 0.2$, which is consistent with
our previous magnetization result. For square lattice infinite-$U$ Hubbard model,
recent DMRG results suggest that phase separation might occur for $\delta>0.2$\cite{DMRGNagaoka}.
On the contrary, for honeycomb lattice infinite-$U$ Hubbard model,  our DMRG results on small clusters don't support phase separation as well as other translational symmetry breaking competing order, e.g., strip order, in the whole region for $\delta<0.25$.  Therefore, we confirm that the observed $p+ip$ SC state is stable for $\delta<0.2$.

\begin{figure}[t]
{\includegraphics[width=0.4\textwidth]{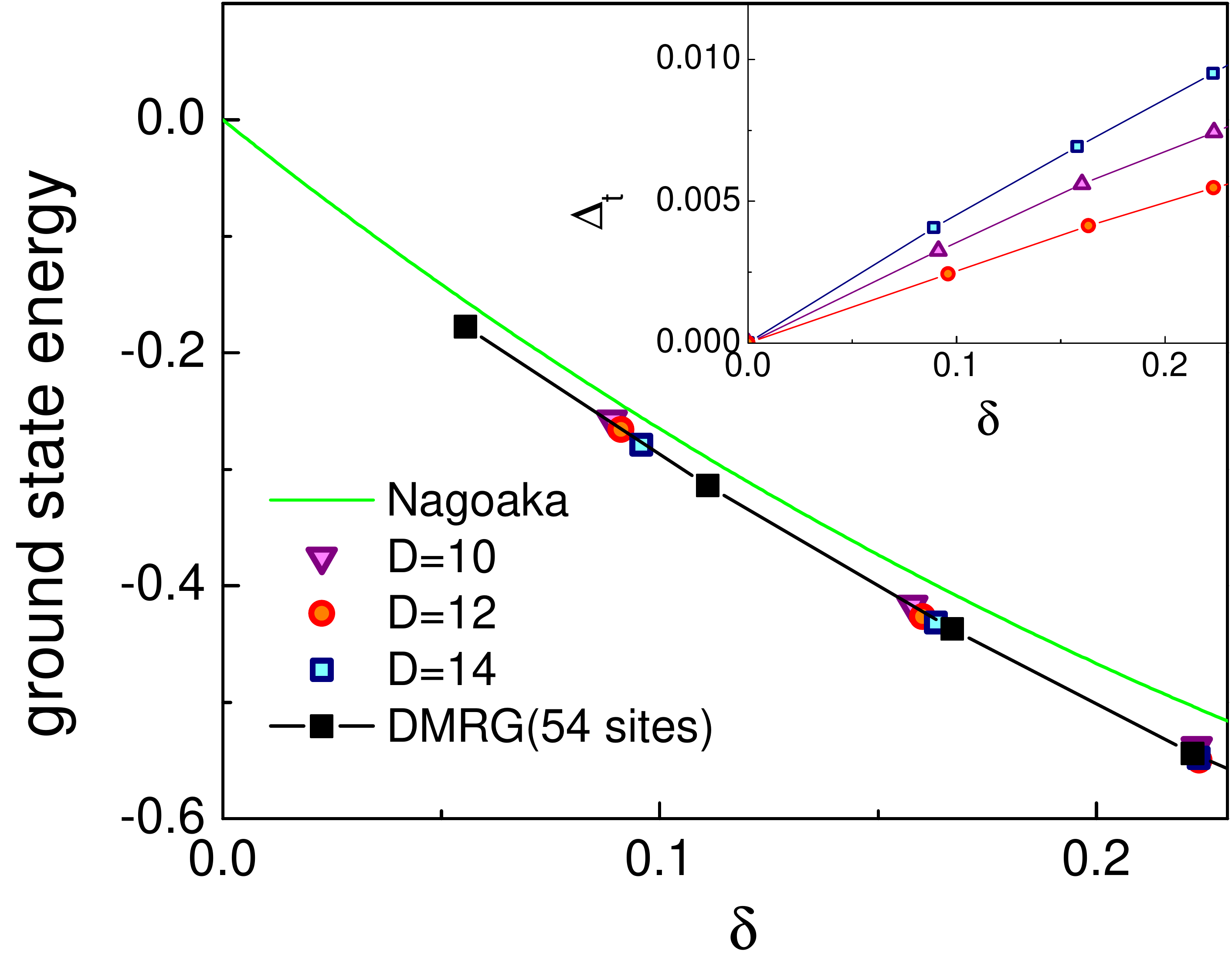}} \caption{(Color online) Ground
state energy as a function of doping for $t-J$ model at $t/J=30$. As a benchmark, we performed
DMRG calculation for a small cluster with $N=54$ sites under PBC.
Insert: $p+ip$ SC order parameter as a function of
doping.}\label{fig:energy50}
\end{figure}

\emph{Instability and stability of the $p+ip$ SC state} -- Although the infinite-$U$ Hubbard model is interesting theoretically, in realistic materials, $U$ is always finite. To investigate the instability of $p+ip$ SC state against a small $J$, we perform both GTPS calculation and DMRG calculation with $t/J=30$. For inner dimension $D=10,12,14$, we
find a systematical consistency for ground energy with both methods for $\delta < 0.25$.\footnote{Again, in the GTERG algorithm, we keep $D_{cut}=160$ for $D=10,12$ and $D_{cut}=196$ for $D=14$ to ensure that the relative energy error is less than $10^{-3}$.}
As seen in Fig. \ref{fig:energy50}, we find the ground state energy is lower than HMS, which clearly shows the instability of HMS. Furthermore, we find that the FM magnetization $m$ vanishes for all $\delta<0.25$, which is quite different from the infinite-$U$ case.

To understand the nature of the ground state phase for large but finite $U$, we further measure the triplet SC order parameter. As seen in the insert of Fig. \ref{fig:energy50}, we find that $\Delta_t$ monotonically decreases with the increase of  $D$, indicating that the $p+ip$ SC order parameter vanishes in $D\rightarrow\infty$ limit. We also measure the AF magnetization and find it vanishes for $\delta>0.1$. All these results suggest a new quantum phase emerges for $\delta>0.1$ with $t/J=30$. We will develop an effective field theory to understand this new phase later; for now, we ask how can we stabilize the $p+ip$ SC state at large but finite $U$.

\begin{figure}[t]
{\includegraphics[width=0.4\textwidth]{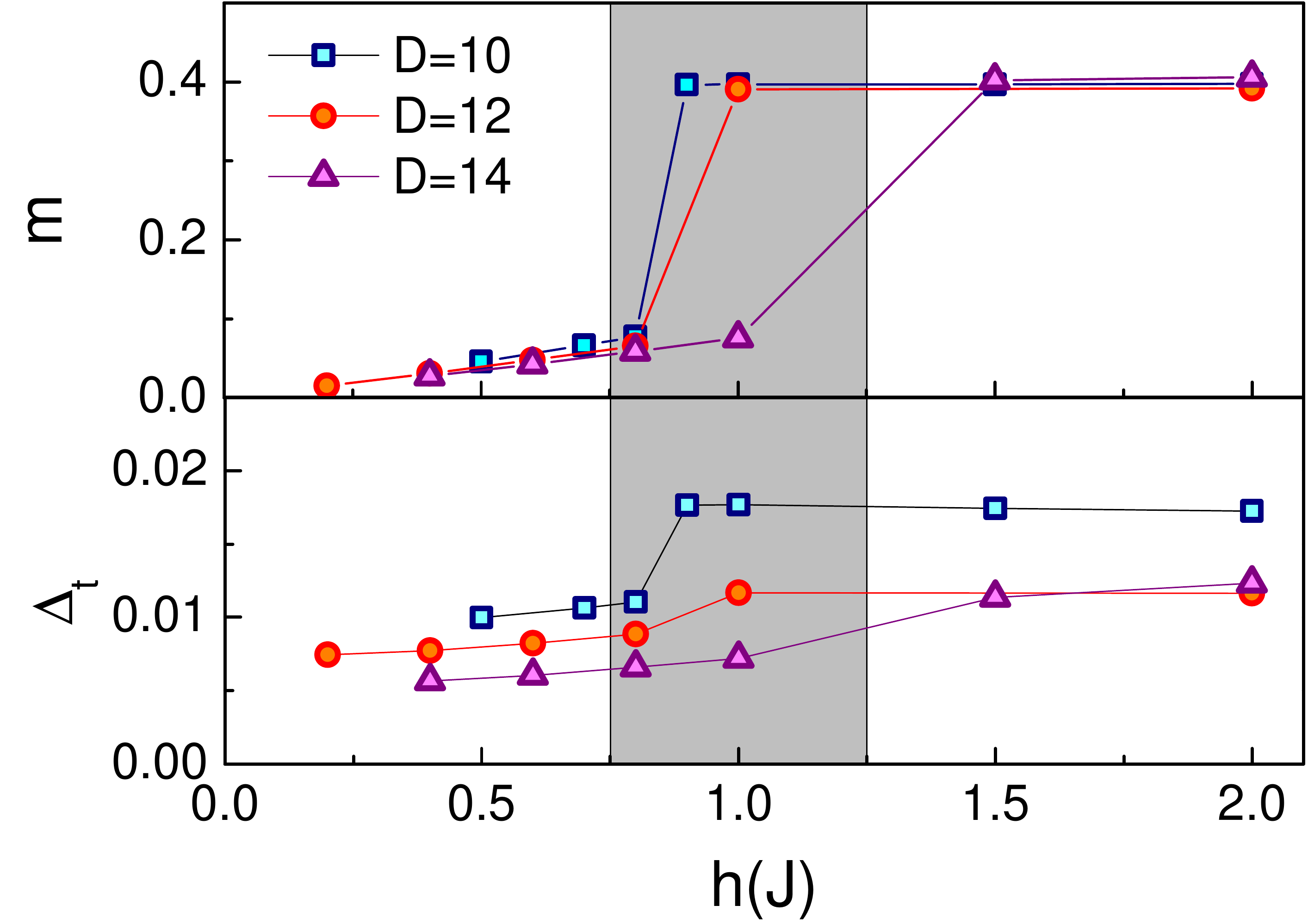}} \caption{(Color online)FM magnetization $m$ and triplet SC order
parameters $\Delta_t$ as a function of Zeeman field for $\delta \sim 0.2$.}\label{fig:field}
\end{figure}

Since the observed $p+ip$ SC state in the infinite-$U$ limit always coexists with an FM ordered state, it is not surprising that a Zeeman field term might stabilize the $p+ip$ SC state. We perform the GTPS calculation for $\delta \sim 0.2$ with a Zeeman field term, e.g., $h S_z$, to confirm this. As seen in Fig. \ref{fig:field}, a phase transition clearly occurs around $h\sim J$. The FM magnetization $m$ suddenly jumps to $m/m_{Max}\sim 0.98$ for $h > J$. For $h<J$, the triplet SC order parameter $\Delta_t$ shows a very strong $D$ dependence, again implying that the $p+ip$ SC order vanishes in the $D\rightarrow\infty$ limit. On the contrary, for $h >J$, $\Delta_t$ saturates at large $D$, e.g., $D=12,14$, which is very similar to the infinite-$U$ case at low doping and indicates the emergence of $p+ip$ SC order. Thus, we conclude that at finite $U$, a Zeeman field of order
$J$ is sufficient to stabilize the $p+ip$ SC state. In realistic materials, the Zeeman field term can be realized by applying in-plane magnetic field.

\emph{Spin-charge separation and skyrmion current mediated pairing} -- Now the central question would be: what are the key mechanisms for the intriguing $p+ip$ SC state. As the numerically observed FM magnetization in $p+ip$ SC state is extremely close to the fully polarized HMS, it is very natural to assume HMS as the "normal" state. Thus we can define the "condensation energy" via $\Delta E=E_{\rm{HMS}}-E_0$($E_0$ is the true ground state). As seen in the insert of Fig. \ref{fig:SC}, the $54$ sites DMRG results(which are consistent with GTPS results)show an approximately linear increase of $\Delta E$ as function of doping. For $\delta <0.15$, we can approximately fit $\Delta E \sim 0.1 \delta t$. In addition, the $p+ip$ SC ordering parameter also increases linearly and can be approximately fitted as $\Delta_t \sim 0.07\delta $. All these observations strongly support the novel phenomenon proposed by P.W.Anderson\cite{RVB}: spin-charge separation!

To construct an effective field theory, we begin with the Schwinger boson representation and decompose the electron operator as $c_{i\sigma}=f_i^\dagger b_{i\sigma}$, where $f_i$ is the fermionic holon while $b_{i\sigma}$ is the bosonic spinon, satisfying the no-double-occupancy constraint $f_i^\dagger f_i+\sum_{\sigma}b_{i\sigma}^\dagger b_{i\sigma}=1$. Inside the HMS phase, we have $b_\uparrow=\sqrt{\rho_0+\delta\rho}e^{i\theta}$ and $b_\downarrow=0$.(We assume a small Zeeman field along $\up$ direction is applied such that the global spin rotational symmetry is broken down to $U(1)$.) After integrating out the quadratic fluctuations of $\delta\rho$, we can use an $XY$ model to describe the spin dynamics. Due to the no-double-occupancy constraint, we also need to introduce an emergent $U(1)$ gauge field $A_\mu$ to couple both spin and charge degrees of freedom:
\begin{eqnarray}
\mathcal{L}_\text{eff} &=& \overline{\psi}_a\gamma^\mu(\partial_\mu-iA_\mu) \psi_a -\mu \overline{\psi}_a\gamma^0\psi_a+\frac{\rho_0}{g}({\partial_\mu\theta}
-A_\mu)^2, \nonumber\\
\end{eqnarray}
where $\psi_a$ is the two component Dirac spinor, $\mu$ is the chemical potential and $a=1,2$ labels two Dirac cones at different momentum points. Summation over repeated indices is everywhere understood.

To understand the instability of the above action, we apply duality transformation and introduce the vortex degrees of freedom(see method section for details):
\begin{eqnarray}
\mathcal{L}_\text{eff} &=& \overline{\psi}_a\gamma^\mu(\partial_\mu-iA_\mu) \psi_a -\mu \overline{\psi}_a\gamma^0\psi_a+\frac{i}{2\pi}\varepsilon^{\mu\nu\lambda}A_{\mu}\partial_\nu a_{\lambda}\nonumber\\&+&\frac{g}{16\pi^2\rho_0}(\varepsilon^{\mu\nu\lambda}\partial_\nu a_{\lambda})^2+|(\partial_\mu+i a_\mu)\phi|^2+m^2|\phi|^2,\nonumber\\\label{effective}
\end{eqnarray}
Here the complex scalar field $\phi$ describes the vortex/anti-vortex field which couples to the dual gauge field $a_\mu$.

Naively, the above action is stable and will not induce attractive interactions for holons, nevertheless, in the dense vortex limit, an FM vortex current -- holon/charge current (non-minimal) coupling term with the form $\mathcal{L}_\text{CC}= j_\mu\varepsilon^{\mu\nu\lambda}\partial_\nu a_{\lambda}$ could arise, where $j_\mu=i [\phi^*(\partial_\mu-i a_\mu)\phi-\phi(\partial_\mu-i a_\mu)\phi^* ]$ is the (gauge invariant) vortex current.(After we integrate out the gauge field $A_\mu$, we will find $\varepsilon^{\mu\nu\lambda}\partial_\nu a_\lambda=2\pi\overline{\psi}_a\gamma^\mu \psi_a$, which is indeed the holon/charge current.)

By adding $\mathcal{L}_\text{CC}$ into $\mathcal{L}_\text{eff}$ and integrating out the vortex field $\phi$ and the gauge field $A_\mu$, we will end up with an effective action for holons. We find that in addition to the usual onsite repulsive interacting term $(\overline{\psi}_a\gamma^\mu \psi_a)^2$ induced by the screened gauge field $A_\mu$, another term with the form $\left[\partial_\mu(\overline{\psi}_a\gamma^\nu \psi_a)\right]^2$ arises.
%\begin{eqnarray}
%&&\mathcal{L}_\text{eff} = \overline{\psi}_a\gamma^\mu \partial_\mu \psi_a -\mu \overline{\psi}_a\gamma^0\psi_a\\ %&&+(\frac{g}{4\rho_0}+\frac{4\pi^2}{m})(\overline{\psi}_a\gamma^\mu \psi_a)^2+\frac{4\pi^2}{m}\left[\partial_\mu(\overline{\psi}_a\gamma^\nu %\psi_a)\right]^2+\cdots,\nonumber\label{effective}
%\end{eqnarray}
Interestingly, it turns out that $\left[\partial_\mu(\overline{\psi}_a\gamma^\nu \psi_a)\right]^2$ is attractive\footnote{We note that the $\mu=\nu$ components vanish due to the constraint $\varepsilon^{\mu\nu\lambda}\partial_\nu a_\lambda=2\pi\overline{\psi}_a\gamma^\mu \psi_a$(since $\varepsilon^{\mu\nu\lambda}\partial_\mu\partial_\nu a_\lambda=0$), and the $\mu\neq \nu$ components lead to attractive interactions. For example, the lattice version of the component $\left[\partial_x(\overline{\psi}\gamma^0 \psi)\right]^2$ for a single Dirac field takes a form $(n_i-n_{i+\v x})^2=-2n_in_{i+\v x}+n_i+n_{i+\v x}$, which is indeed an attractive interaction(up to a chemical potential term). Apparently, all the other $\mu\neq\nu$ components are also attractive due to the Lorentz invariance of $\left[\partial_\mu(\overline{\psi}\gamma^\nu \psi)\right]^2$.} and will dominate at low energy. Therefore, this term will lead to the instability of holon Fermi surface. It is well known that the spinless fermion with attractive interactions must have a $p+ip$ pairing symmetry, thus we find the key mechanism(pairing force) for the numerically observed $p+ip$ SC state in honeycomb lattice infinite-$U$ Hubbard model.

On the other hand, when $m_\zeta^2<0$,
\footnote{In general, a $\lambda |\phi|^4$ term need to be added into $\mathcal{L}_\text{eff}$, which comes from the fact that more than one segments of vortices cannot occupy the same place.}
vortex condensation happens and a mass term of $a_\mu$ is induced, and it generates a Maxwell term for $A_\mu$.
\begin{eqnarray}
\mathcal{L}_\text{eff}^{\rm{NF}} &=& \overline{\psi}_a\gamma^\mu(\partial_\mu-iA_\mu) \psi_a -\mu \overline{\psi}_a\gamma^0\psi_a+\frac{1}{g^\prime}(\varepsilon^{\mu\nu\lambda}\partial_\nu A_{\lambda})^2,\nonumber\\
\end{eqnarray}
We propose the above action as the effective field theory description for the potential non-Fermi liquid phase arising at small but finite $J$ for $\delta>0.1$.

Finally, we provide a more straightforward way to understand the origin of charge-current and FM skyrmion(vortex) current coupling in the infinite-$U$ Hubbard model. Indeed, with reference to fully up spin polarized sector, the infinite-$U$ Hubbard model can be rewritten as (see method section for detailed derivations):
\begin{eqnarray}
H  & = &  t(1-\delta_0)^2 \sum_{\langle ij \rangle}( c^\dagger_{i\up} c_{j\up} + h.c.)  \nonumber \\
&+&  \frac{1}{2} t\sum_{\langle ij \rangle} ( c^\dagger_{i\up} c_{j\up} +  c^\dagger_{j\up} c_{i\up})  (S^-_i S^+_j  + S^-_j S^+_i ) \nonumber \\
&+&   \frac{1}{2} t\sum_{\langle ij \rangle} ( c^\dagger_{i\up} c_{j\up} -  c^\dagger_{j\up} c_{i\up})  (S^-_i S^+_j  - S^-_j S^+_i ),\label{newH}
\end{eqnarray}
where $S_i^-\equiv c_{i\down}^\dagger c_{i\up}$ and $S_j^+\equiv c_{j\up}^\dagger c_{i\down}$ are the spin flipping operators. There is a local constraint that prevents any site from having an up spin hole and a reversed Pauli spin; this makes the above representation useful only when $\langle n_i^\down\rangle=\delta_0 \ll 1$.
Spin charge coupling lurking in the infinite-$U$ Hubbard model is manifested in the above representation.  The charge current is coupled to a spin nematic chiral current represented by ${\vec z}\cdot ({\vec S}_i \times{\vec S}_j) \equiv (S^-_i S^+_j  - S^-_j S^+_i) $. We can also view this as a topological spin current density or a scalar chirality, ${\vec m}\cdot ({\vec S}_i \times{\vec S}_j)$, with the global magnetization vector ${\vec m}$  pointing along $\up$-direction and two spins are at site $i$ and $j$, which is indeed the FM skyrmion(vortex) current. If we add a Gaussian term $\frac{1}{\kappa}({\vec z}\cdot ({\vec S}_i \times{\vec S}_j)^2$ to FM skyrmion(vortex) current, which is regarded as independent low energy degrees of freedom here, a nearest neighbor(NN) attractive interaction with the form $- 2 t^2 \kappa n_{i\up} n_{j\up}$ will arise after we integrate out the quantum fluctuations of FM skyrmion (vortex) current.

In conclusion, we report the theoretical discovery of a $p+ip$ wave
superconducting ground state in honeycomb lattice infinite-$U$ Hubbard model. It would be interesting to search for this physics
in experiment. The recently discovered spin $1/2$ honeycomb lattice
Mott-insulator InV$_{1/3}$Cu$_{2/3}$O$_3$\cite{InVCuO} would be an
appealing candidate if it could be doped experimentally. It will also be important to perform new experiments and look at existing results on mono layer $^3$He adsorbed on substrate\cite{3He}, which provide a nearly perfect avenue for the realization of infinite-$U$ Hubbard model on a few lattice geometries, including the honeycomb lattice. The recently discovered 2D ferromagnetism in conducting organic
 2D layer grown on graphene\cite{organic} provides another important playground
 to test our ideas of $p+ip$ superconductivity.

%This novel quantum effect presented in this article seems to be unique to itinerant ferromagnetism. It is likely to have implications for well known, but still less understood itinerant ferromagnets such as Fe, Co and Ni, in its bulk form or few layer thick 2 dimensional form.

\section{Method}
\begin{figure}[h]
{\includegraphics[width=0.3\textwidth]{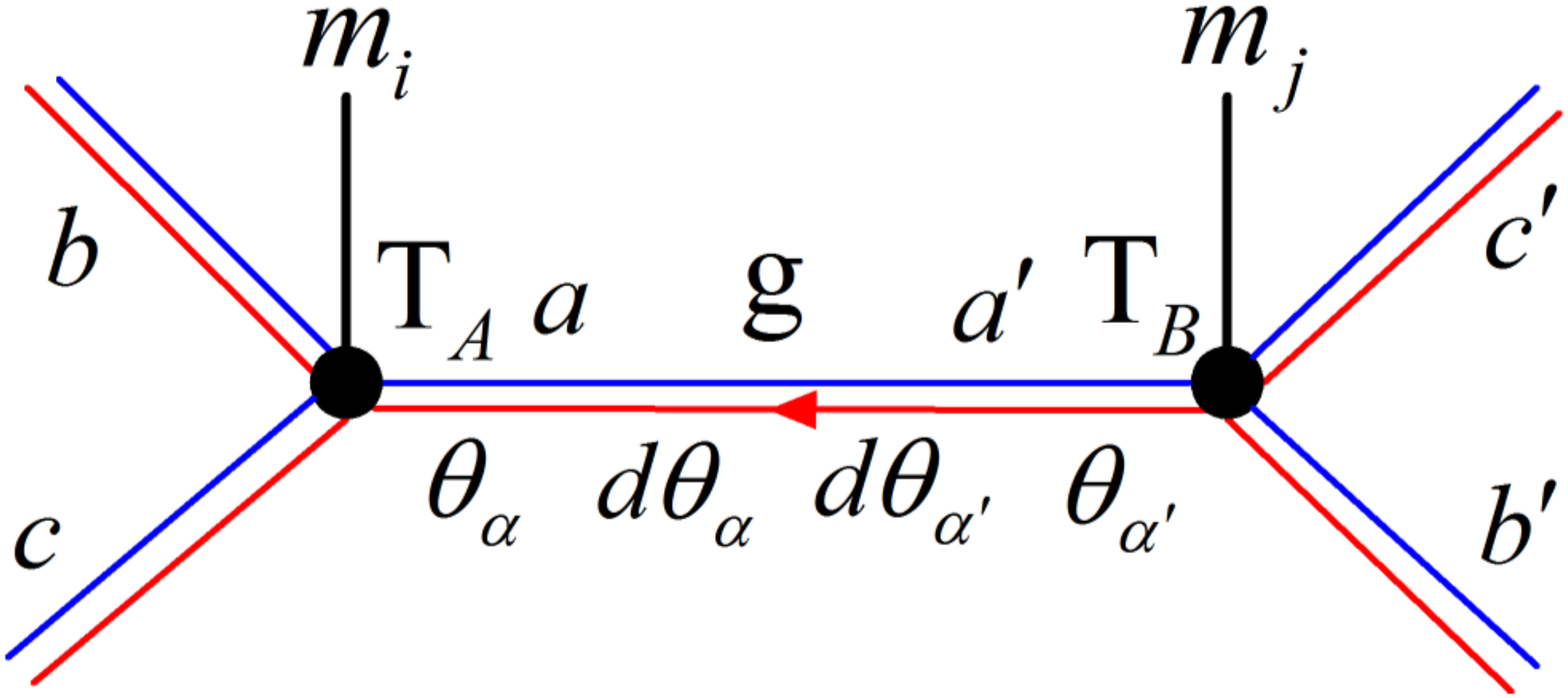}}
\caption{(Color online)Graphic representation of the GTPS on a honeycomb lattice.
$\textbf{T}_A$ and $\textbf{T}_B$, which contain $\theta$ are
defined on the sublattices $A$ and $B$ for each unit cell. The
Grassmann metric $\textbf{g}$ containing $\dd \theta$ is defined on
the links that connect the Grassmann tensors $\textbf{T}_A$ and
$\textbf{T}_B$. The blue lines represent the fermion parity even
indices while the red lines represent the fermion parity odd indices
of the virtual states. Notice an arrow from $A$ to $B$ represents
the ordering convention $\dd \theta_\alpha\dd
\theta_{\alpha^\prime}$ that we use for the Grassmann
metric.}\label{fig:GrassmannTPS}
\end{figure}

{\bf Grassmann tensor product state} --
We use the standard form of GTPS
as our variational wavefunction. We further assume a translationally
invariant ansatz, and thus it is specified by just two different
Grassmann tensors $\textbf{T}_A,\textbf{T}_B$ on sublattices $A,B$ of
each unit cell:
\begin{eqnarray}
&& \Psi(\{m_i\},\{m_j \}) \\
\nonumber &&  = {\rm{tTr}} \int \prod_{\langle ij\rangle}\textbf{g}_{a
a^\prime} \prod_{i\in A} {\textbf{T}}^{m_i}_{A;abc} \prod_{j\in B}
{\textbf{T}}^{m_j}_{B;a^\prime b^\prime c^\prime},\label{GTPS}
\end{eqnarray}
with
\begin{eqnarray}
{\textbf{T}}^{m_{i}}_{A;abc}&=& {T}^{m_{i}}_{A;abc}
\theta_\alpha^{P^f(a)} \theta_\beta^{P^f(b)}
\theta_\gamma^{P^f(c)},  \nonumber\\
 {\textbf{T}}^{m_{j}}_{B,a^\prime b^\prime c^\prime}&=&
{T}^{m_{j}}_{B;a^\prime b^\prime c^\prime}
\theta_{\alpha^\prime}^{P^f(a^\prime)}
\theta_{\beta^\prime}^{P^f(b^\prime)}
\theta_{\gamma^\prime}^{P^f(c^\prime)},    \nonumber\\
\textbf{g}_{aa^\prime}&=& \delta_{aa^\prime}{\dd
\theta}_\alpha^{P^f(a)} {\dd
\theta}_{\alpha^\prime}^{P^f(a^\prime)}.
\end{eqnarray}
We notice that the symbol $\rm{tTr}$ means tensor contraction of the
inner indices $\{a\}$. Here
$\theta_{\alpha(\beta,\gamma)},\mathrm{d}\theta_{\alpha(\beta,\gamma)}$
are the Grassmann numbers and dual Grassmann numbers respectively
defined on the link $a(b,c)$and they satisfy the Grassmann algebra:
\begin{align}
 \theta_\alpha\theta_\beta&=-\theta_\beta\theta_\alpha,
&
 \dd{\theta_\alpha}\dd{\theta_\beta}&=-\dd{\theta_\beta}\dd{\theta_\alpha},
\nonumber\\
\int \dd{\theta_\alpha}\theta_\beta &=\delta_{\alpha\beta} & \int
\dd{\theta_\alpha} 1&=0 .
\end{align}
As shown in Fig.\ref{fig:GrassmannTPS}, $a,b,c=1,2,\ldots,D$ are the
virtual indices carrying a fermion parity $P^f(a)=0,1$. In this
paper, we choose $D$ to be even and assume there are \emph{equal}
numbers of fermion parity even/odd indices, which might be not
necessary in general. Those indices with odd parity are always
associated with a Grassmann number on the corresponding link and the
metric $\textbf{g}_{aa^\prime}$ is the Grassmann generalization of
the canonical delta function. The complex coefficients
${T}^{m_{i}}_{A;abc}$ and ${T}^{m_{j}}_{B;a^\prime b^\prime
c^\prime}$ are the variational parameters.

Notice that $m_{i}$ is the physical index of the $t$-$J$ model on
site $i$, which can take three different values, $o,\uparrow$ and
$\downarrow$, representing the hole, spin-up electron and spin-down
electron states.   We choose the hole representation in our
calculations, and thus the hole state has an odd parity $P^f(o)=1$
while the electron states have even parity
$P^f(\uparrow,\downarrow)=0$. On each site, the non-zero components
of the Grassmann tensors should satisfy the parity conservation
constraint:
\begin{align}
P^f(m_{i})+P^f(a)+P^f(b)+P^f(c)=0 ({\rm mod}\, 2) .
\end{align}
Since the wavefunction Eq.(\ref{GTPS}) does not have a definite
fermion number, we use the grand canonical ensemble, adding a
chemical potential term to control the average hole
concentration.

{\bf Dual vortex representation of Higgs phase} --
To understand the pairing force for $p+ip$ SC order, we first perform the Hubbard-Stratonovich transformation to decouple the $({\partial_\mu\theta}
-A_\mu)^2$ term and decompose the $\theta$ field as $\theta=\eta+\zeta$ where $\eta$ is the smooth part of the phase fluctuation and $\zeta$ is the vortex part.
\begin{eqnarray}
  \mathcal{L}_\text{eff}&=& \overline{\psi}_a\gamma^\mu(\partial_\mu-iA_\mu) \psi_a -\mu \overline{\psi}_a\gamma^0\psi_a\nonumber\\&+&\frac{g}{4\rho_0}(\xi^\mu)^2-i\xi^\mu(  {\partial_\mu\eta}+{\partial_\mu\zeta}-A_\mu)  ,\label{effective}
\end{eqnarray}
By integrating the smooth fluctuations $\eta$, we obtain the constraint $\partial_\mu\xi^\mu=0$. Resolving the constraint leads to $\xi^\mu=\frac{1}{2\pi}\varepsilon^{\mu\nu\lambda}\partial_\nu a_{\lambda}$.(We note that the dual $U(1)$ gauge field $a_\mu$ is non-compact.)
\begin{eqnarray}
  \mathcal{L}_\text{eff}&=& \overline{\psi}_a\gamma^\mu(\partial_\mu-iA_\mu) \psi_a -\mu \overline{\psi}_a\gamma^0\psi_a+\frac{i}{2\pi}\varepsilon^{\mu\nu\lambda}A_{\mu}\partial_\nu a_{\lambda}\nonumber\\& +&\frac{g}{16\pi^2\rho_0}(\varepsilon^{\mu\nu\lambda}\partial_\nu a_{\lambda})^2+ i a_\mu j_\mu+ A \sum_i \int ds_i   ,
\end{eqnarray}
where $j_\mu=\frac{1}{2\pi}\varepsilon^{\mu\nu\lambda}\partial_\nu \partial_\lambda \zeta$ is the vortex current. We also add a term
$A \sum_i \int ds_i$ which corresponds to the energy cost at the core of the vortex, where $A$ is the energy cost per unit length and $\in ds_i$ is the length of the world line of $i$th vortex. Since summation over vortex line configurations can be reexpressed by integrating over a free massive complex field $\phi$\cite{Nagaosa}, we end up with a dual action evolving vortex degrees of freedoms $\phi$ as well as the dual gauge field $a_\mu$:
\begin{eqnarray}
  \mathcal{L}_\text{eff}&=& \overline{\psi}_a\gamma^\mu(\partial_\mu-iA_\mu) \psi_a -\mu \overline{\psi}_a\gamma^0\psi_a+\frac{i}{2\pi}\varepsilon^{\mu\nu\lambda}A_{\mu}\partial_\nu a_{\lambda}\nonumber\\ &+&\frac{g}{16\pi^2\rho_0}(\varepsilon^{\mu\nu\lambda}\partial_\nu a_{\lambda})^2 +|(\partial_\mu+i a_\mu)\phi|^2+m^2|\phi|^2,\nonumber\\
\end{eqnarray}

{\bf Charge current -- topological spin current interaction}
Consider the infinite-$U$ Hubbard model assuming we are staying in almost fully polarized up spin sectors with a small density of down spins, $\langle n_i^\down\rangle=\delta_0 \ll 1$. The infinite-$U$ Hubbard model in these sectors can be exactly rewritten as:

\begin{eqnarray}
H &=& t\sum_{\langle ij\rangle}(1-n_{i\down})c_{i\up}^\dagger c_{j\up}(1-n_{j\down})+h.c.\\&+&
 t\sum_{\langle ij\rangle}(1-n_{i\up})c_{i\down}^\dagger c_{j\down}(1-n_{j\up})+h.c.\nonumber\\
 & = & t(1-\delta_0)^2 \sum_{\langle ij\rangle} c_{i\up}^\dagger c_{j\up}+h.c.
 + t\sum_{\langle ij\rangle}c_{i\up}^\dagger c_{j\up} S_i^- S_j^+ +h.c.\nonumber
\end{eqnarray}

We note that $(1-n_{i\up})c_{i\down}^\dagger=c_{i\up}c_{i\up}^\dagger c_{i\down}^\dagger=S_i^- c_{i\up}^\dagger$, where $S_i^-\equiv c_{i\down}^\dagger c_{i\up}$. Similarly, $c_{j\down}(1-n_{j\up})=c_{j\down}c_{j\up} c_{j\up}^\dagger= c_{j\up} S_j^+$. An onsite double occupancy constraint between up spin hole and a down Pauli spin can be well approximated by
replacing hopping $t$ by $t(1-\delta_0)^2$ for the up spin electron hopping term.

The first term in the above is the usual kinetic term for polarized electron while the second term is a backflow of down spins. Physics of this important term becomes more transparent when we rewrite the hopping term in terms of kinetic energy and charge current operators as:

\begin{eqnarray}
H  & = &   t(1-\delta_0)^2 \sum_{\langle ij \rangle}( c^\dagger_{i\up} c_{j\up} + h.c.)  \nonumber \\
&+&  \frac{1}{2}  t\sum_{\langle ij \rangle} ( c^\dagger_{i\up} c_{j\up} +  c^\dagger_{j\up} c_{i\up})  (S^-_i S^+_j  + S^-_j S^+_i ) \nonumber \\
&+&   \frac{1}{2} t\sum_{\langle ij \rangle} ( c^\dagger_{i\up} c_{j\up} -  c^\dagger_{j\up} c_{i\up})  (S^-_i S^+_j  - S^-_j S^+_i ),\label{newH}
\end{eqnarray}

We identify $(S^-_i S^+_j  + S^-_j S^+_i ) \equiv (S^i_x S^j_x + S^i_y S^j_y)$ with $XY$ spin-spin interaction in the plane perpendicular to the direction of FM polarization. It is easily seen that this coupling between kinetic energy of up spin holes and down spin electron encourages an antiferromagnetic spin correlation in the transverse direction. Similarly we identify
$(S^-_i S^+_j-S^-_j S^+_i ) \equiv {\vec z}\cdot ({\vec S}_i \times{\vec S}_j)$ with scalar spin chirality, with reference to the FM moment vector. This term encourages coupling between Skyrmion current and charge current. It is the gain in energy arising from this term that favors a finite density of Skyrmions in the ground state. In other words, the dynamic holes stir the skyrmion gas and in the process undergo a pairing instability in the $p+ip$ channel. 

%(please add this note as a reference in an apropriate place): Using a different argument Kumar (Brijesh Kumar) arrives at an elegent spin rotataional invariant representation of infinite-U Hubbard model (Brijesh Kumar), interms of Pauli spin opertors and spinless fermions. However manifest spin rotatinal invariance renders structure of the Hamiltonian somwhat involved and hides the physics we are after.

\section*{Acknowledgment}
Z.C.G would like to thank Leon Balents, Dong-Ning Sheng and Patrick Lee for valuable discussions. This work is supported by the Government of Canada through Industry Canada and by the Province of Ontario through the Ministry of Research and Innovation. H.C.J was supported by the U.S. Department of Energy. G.B acknowledges Science and Engineering Research Board, India as a SERB Distinguished Fellow.

\end{document}